# INITIAL GROWTH OF TIN ON NIOBIUM FOR VAPOR DIFFUSION COATING OF Nb$_3$Sn

U. Pudasaini[1], G. Eremeev[2], C. E. Reece[2], J. Tuggle[3], M. J. Kelley[1,2,3]

[1]Applied Science Department, The College of William and Mary, Williamsburg, VA 23185, USA
[2]Thomas Jefferson National Accelerator Facility, Newport News, VA 23606, USA
[3]Virginia Polytechnic Institute and State University, Blacksburg, VA 24061, USA

*Abstract*

Nb$_3$Sn has the potential to achieve superior performance in terms of quality factor, accelerating gradient and operating temperature (4.2 K vs 2 K) resulting in significant reduction in both capital and operating costs compared to traditional niobium SRF accelerator cavities. Tin vapor diffusion coating of Nb$_3$Sn on niobium appears to be a simple, yet most efficient technique so far to fabricate such cavities. Here, cavity interior surface coatings are obtained by a two step process: "nucleation" followed by "deposition". The first step is normally accomplished with Sn/SnCl$_2$ at a constant low temperature (~500 °C) for several hours. To elucidate the role of this step, we systematically studied the niobium surface nucleated under varying process conditions. The surfaces obtained in typical tin/tin chloride processes were characterized using SEM/EDS, AFM, XPS, SAM and TEM. Examination of the surfaces nucleated under the "standard" conditions revealed not only tin particles, but also tin film on the surfaces resembling the surface obtained by Stranski-Krastanov growth mode. All the nucleation attempted with SnCl$_2$ yielded better uniformity of Nb$_3$Sn coating compared to coating obtained without nucleation, which often included random patchy regions with irregular grain structure. Even though the variation of nucleation parameters was able to produce different surfaces following nucleation, no evidence was found for any significant impact on the final coating.

## 1    Introduction

Modern particle accelerators use superconducting radio frequency (SRF) cavities to accelerate a beam of charged particles. Niobium has been the material of choice so far to fabricate SRF cavities. However, their operation is very expensive as it needs to maintain a ~2 K cryogenic temperature for effective operation. Further, after more than five decades of research, the performance of niobium cavities is approaching the theoretical limit set by intrinsic material properties. Application of new materials instead can improve the performance as well as reduce the cost of SRF accelerators [1].

Discovered in 1954 by Bernd Matthias [2], Nb$_3$Sn is a promising alternative. Its critical temperature (~ 18 K) and predicted superheating field (~ 400 mT) are nearly twice that of niobium, thereby offering the possibility of attaining higher quality factor and accelerating gradient at any given temperature [3] . It can also allow an increase of the cavity operation temperature, resulting in significant reduction in both capital and operating cost for the cryoplant. However, Nb$_3$Sn is a challenging material for cavity application because of lower thermal conductivity and brittleness. That restricts its application to coating form. SRF cavities typically have complicated geometries that need to be coated uniformly, which limits the coating techniques available.

Several techniques have been attempted to deposit Nb3Sn layers: chemical vapor deposition, co-evaporation, tin bath dipping and annealing, pulse laser deposition, electro-deposition, sputtering etc. [4-11]. However, vapor diffusion coating of Nb3Sn, attributed to Saur and Wurm [12], is the most favourable and successful technique so far. Preparation of Nb3Sn cavities by diffusion coating on niobium cavities dates back to 1970's [13-15]. Several research institutions working currently to develop Nb3Sn coated cavities use it [16-18] . Recent performance results of such cavities are very promising, attaining quality factor >10$^{10}$ operating at 4.2 K with gradient more than 15 MV/m [19,20] . The essence of the process is to transport tin vapor to the niobium substrate, and provide the high temperature environment, >930 ºC, to form the Nb3Sn phase exclusively, which is determined by the binary phase diagram [21]. Early attempts at Siemens AG with this method resulted in regions which were not covered completely with Nb3Sn [22]. The speculated cause was irregular nucleation. They prescribed to anodize the Nb-substrate prior to coating while setting the temperature of tin source higher than the substrate temperature, which they believed enhances the nucleation by increasing the tin vapor supply. Another solution was to add a small amount of Sn halide which can yield an increased Sn supply at lower temperature. The idea was adopted with expected benefit from the higher vapor pressure of tin halide than elemental tin [23]. For example, comparison of Sn vapor pressure to SnCl$_2$ vapor pressure is shown in Figure 2. SnCl$_2$ evaporates at temperature of about 500 ºC to deposit tin sites on the niobium surface by decomposition, which were assumed to act as Nb-Sn nucleation sites.

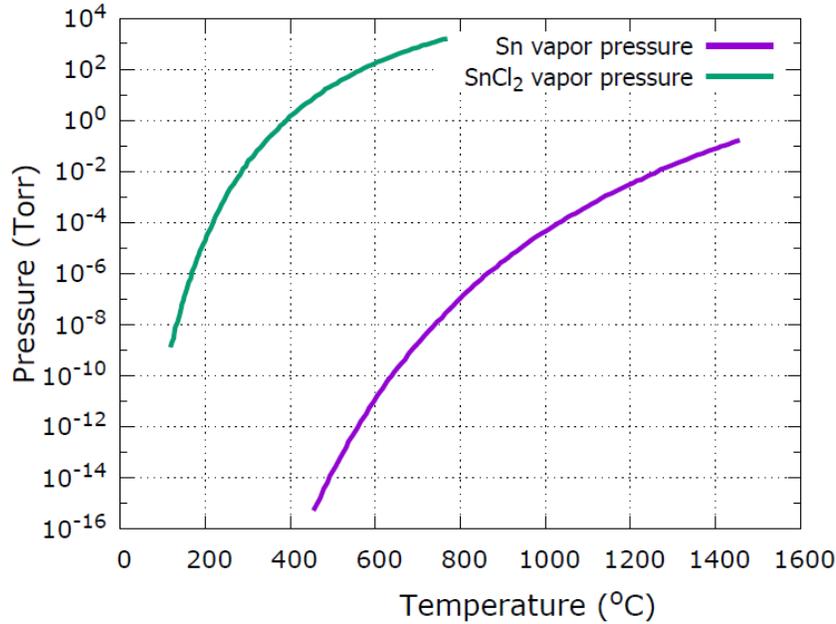

Figure 1: Vapor pressure of tin and tin chloride. The picture is reproduced from [24] which cites [25-27].

Research institutions following a similar Nb$_3$Sn coating recipe later preferred SnCl$_2$, combined sometimes with substrate preanodization or adjusting a temperature gradient between the tin source and the substrate niobium. Later studies suggested that substrate preanodization is not mandatory, and it also results in RRR degradation due to oxygen absorption from the anodic layer [28]. Application of Sn halide is especially helpful for the coating systems that do not have a secondary heater to set up a temperature gradient between the Sn source and the substrate, as discussed above.

A typical Nb3Sn coating process shown in Fig. 1, depicts the process at JLab. The first temperature plateau at 500 ºC is dedicated to the nucleation process whereas the second at 1200 ºC facilitates deposition.

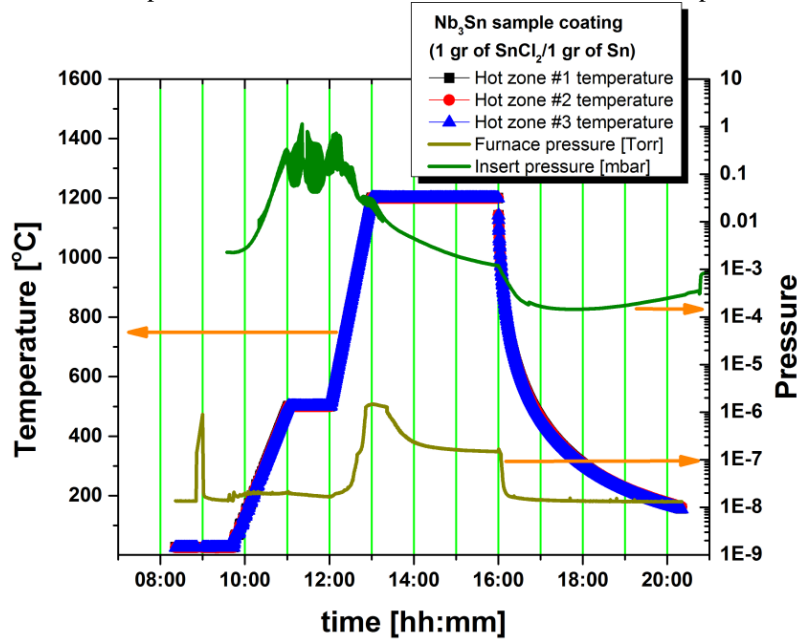

Figure 2: Temperature profile used for coating Nb$_3$Sn on niobium samples at Jlab [29].

While the inclusion of a nucleation step has a long history, only limited research has been done to understand the fundamental atomic process or its relevance and effect on the resulting coating. The purpose of this work was to gain more insight into the process of tin deposition during the nucleation step.

## 2  Experimental Description

### 2.1  Materials

The niobium samples were 10 mm x 10 mm coupons, prepared by EDM cutting from 3 or 4 mm thick, high RRR (~300) sheet material of the type used to fabricate SRF cavities. All were subjected to buffered chemical polishing (BCP) using a solution of 49% HF, 70% $HNO_3$ and 85% $H_3PO_4$ in the ratio of 1:1:1 by volume with minimum removal of 50 μm. These samples further received metallographic polishing, also known as nanopolishing (NP), which typically removes > 100 μm and produces smoother surfaces favourable to most material characterization techniques. The average roughness of NP samples was below 5 nm as measured from 50 μm x 50 μm area scans using atomic force microscopy (AFM) [29].

### 2.2  Experimental Setup

The coating deposition system comprised two parts: the furnace to provide a clean heating environment and the insert, made of niobium, inside which the experimental setup is loaded for each experiment. The temperature was monitored using three thermocouples outside the insert. A detailed description of the coating deposition system can be found in [18].

The insert comprises a sample chamber made of niobium with a shelf to mount coupon samples inside, Fig. 3. One gram of 99.999% or better purity tin shots (~3 mg/cm2 of surface to be coated) and the chosen amount of 99.99% tin chloride powder, both purchased from American Elements, were packaged loosely in niobium foil, and placed on the niobium plate/foil which covered the bottom end of sample chamber. Sn and $SnCl_2$ vapor are expected to exit from the package readily from the narrow openings. The top end was later covered by niobium plate/foil after mounting the experimental coupon samples inside. The sample chamber, samples, chemicals, and the covering foils/plates were assembled in the clean room to limit any contamination before installation into the coating deposition system.

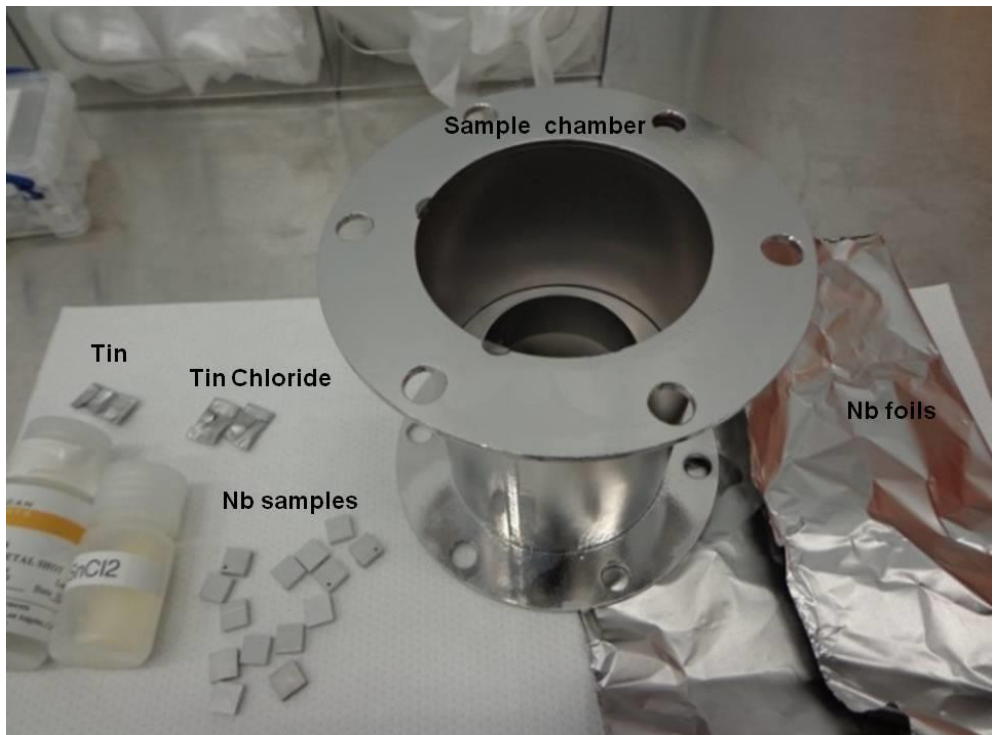

Figure 3: Sample chamber, samples, tin and tin chloride packages, and niobium foils before set up.

*2.3  Experiments*

The insert was pumped down to 10-5 – 10-6 Torr before the heating profile was initiated. The temperature inside the insert was raised at the rate of 6 ºC/min until it reached the target nucleation temperatures as shown in Table 1. These temperatures were maintained constant for different time periods (nucleation time) before ceasing the heating process. After the nucleation at given condition, the chamber was allowed to cool down in vacuum. After reaching to room temperature, the insert was purged with nitrogen to regain atmospheric pressure. The sample chamber was then taken out to remove the nucleated samples. The amount of tin chloride was varied in some experiments.

Table 1: Nucleation Experiments

| Nucleation Temperature | 300 °C | 400 °C | 450 °C | 500 °C |
|---|---|---|---|---|
| Nucleation Time | 1 hour - 4 hour | 1 hour | 1 hour | 5 minute, 5hour |
| Amount of $SnCl_2$ | 1 g | 1 g | 1 g | 5 mg, 1 g |
| Amount of Sn | 1 g | 1 g | 1 g | 0 g, 1 g |

Some experiments featured pre-anodized niobium samples along with regular niobium samples. A selected few nucleation profiles were employed to deposit the complete $Nb_3Sn$ coatings by a coating step as shown in Figure 2. Other experiments included interruption of coating process at different stages with/without nucleation to further understand the effect of nucleation in subsequent coating.

*2.4  Characterization*

An Hitachi 4700 field emission scanning electron microscope (FE-SEM) equipped with an energy dispersive X-ray spectroscopy (EDS) detector was used to examine the nucleated samples. SEM images were taken after each experiment, and elemental composition was analysed with EDS. All SEM images were taken at 0° tilt angle, using a 12/15 kV accelerating voltage.

The topographic examination used a Digital Instruments IV AFM in tapping mode using aluminium reflex coated silicon tips with radius < 10 nm, resonant frequency 190 KHz and force constant 48 N/m. The selected samples were scanned for 5 µm x 5 µm and 1 µm x 1 µm sizes with 512 x 512 data points. The expected lateral resolution of AFM is defined by the spacing of sampling points. For example, the expected lateral resolution for 5 µm scan is 5 µm / 512 (~10 nm) at optimal conditions.

Surface sensitive elemental analysis was done using X-ray photoelectron spectroscopy (XPS), probing the elemental composition of the first few nanometers of surfaces down to roughly a part per thousand. The XPS measurements were carried out in an ULVAC-PHI "Quantera SXM" instrument equipped with a monochromated Al Kα X-ray source. Surface analyses were collected at 50 W/15 kV using a 200 μm spot size, 45° take off angle.

The scanning Auger microscopy (SAM) PHI 680 system was employed for certain samples. It consists FE-SEM with a Schottky emission cathode, a secondary electron detector, and an axial cylindrical mirror analyzer with a multi-channel detector to collect Auger electrons produced during electron imaging. Very small spot sizes can be realized with this instrument, down to 7 nm.

Transmission electron microscopy (TEM) imaging of the cross-section of nucleated sample was carried out using JEOL 2100 operating at 200 kV. The specimen was prepared by focused ion beam (FIB) sectioning by utilizing Helios Nano Lab 600 using the lift-out technique. In order to preserve the surface of nucleated sample and create an intact cross section, a protective layer of Pt was deposited on the sample surface over the area of interest prior sectioning. Initial material removal steps were performed at the highest removal rate of 30 kV. The final polishing step was done with 2kV Ga ion at an angle of 7 degrees.

# 3 Results and Discussion

## 3.1 *Nucleation Temperature Variation*

Temperatures of 300 ℃, 400 ℃, 450 ℃ and 500 ℃ were chosen to study the effect of nucleation temperature. Since the vapor pressure of $SnCl_2$ drops rapidly below 300 ℃, see Fig. 1, this was chosen as the lower bound of the studied temperature range. The nucleation time was kept constant at one hour, and the amount of $SnCl_2$ was fixed to 1 g for each experiment. Residual $SnCl_2$ was found inside the niobium foil containing supplied $SnCl_2$, only after 300 ℃ and 400 ℃.

The chosen nucleation processes were found to produce different surfaces. After 300 ℃ nucleation, a significant portion of $SnCl_2$ was found inside each package. SEM images revealed nanometer-sized particles, assumed to be tin, on the niobium surface, Fig. 4(a). However, only niobium was detected with EDS at 15 kV, possibly due to the shallow coverage of very small amount of deposited tin.

The nucleation temperature 400 ℃ produced distinct features on the surface, Fig 4(b). Besides micron sized spherical particles, extended interconnected 'mud-crack' like features were seen in SEM images. Several SEM images, not shown here, inferred that the big particles were formed by the accumulation of those features. Both niobium and tin signals were detected with EDS not only at the spherical particles, which we assume are tin, but also between the particles, where the irregular structures can be seen in SEM image.

Nucleation temperature of 450 ℃ and 500 ℃ created similar looking surfaces with visible sub-micron sized spherical particles under SEM, appearing as bright features in Fig 4(c) and 4(d). The size and distribution of tin particles was slightly different between these two temperatures. Spherical particles were examined with EDS, showed 40-70 at. % Sn depending on size. Observed tin amount from these particles strongly suggested them to be tin particles. EDS was unable to detect any tin in between those particles. Estimated surface coverage of the particles from SEM images was always less than 10%.

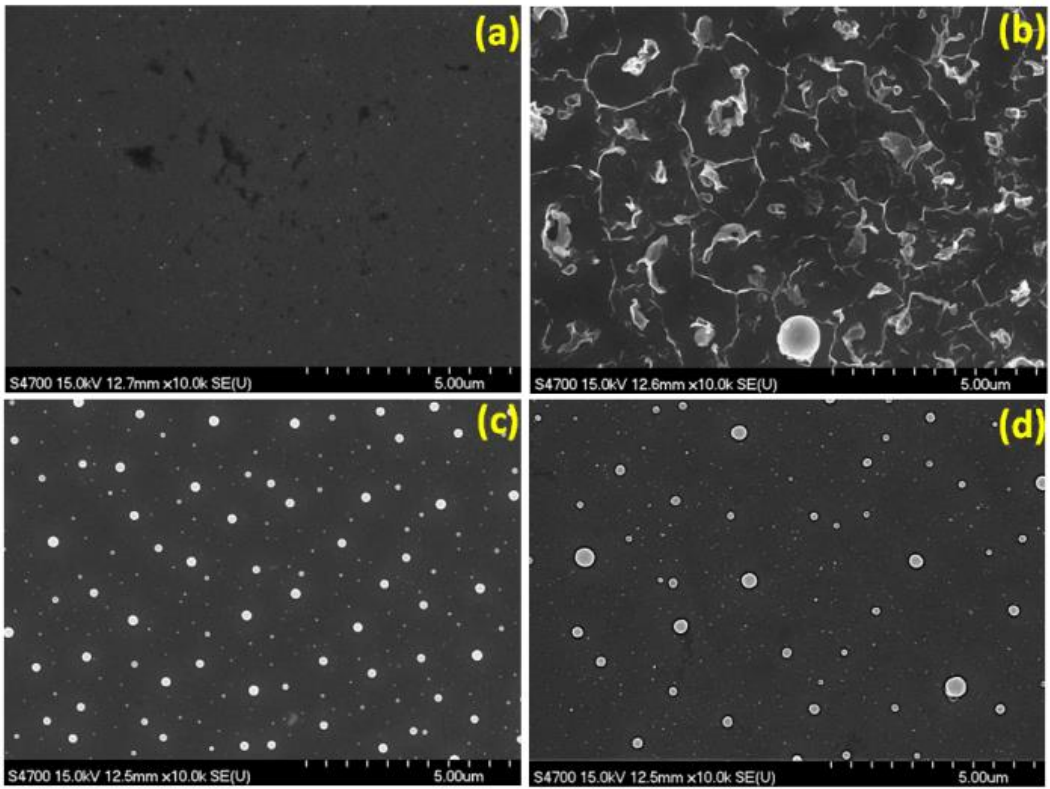

Figure 4: SEM images obtained from samples activated at nucleation temperature of (a) 300 ℃, (b) 400 ℃, (c) 450 ℃, and (d) 500 ℃ with constant nucleation time of an hour. Circular bright features are tin particles when examined with EDS [30].

Since lateral and depth resolution of EDS is on the order of a micron, which is greater than the size of particles observed after 450 ºC or 500 ºC nucleations, samples from each experiment were examined using XPS. Table 1 shows the XPS analysis of representative samples from each experiment. Carbon and oxygen were normally found, which is expected from post-experiment exposure to ambient atmosphere. The samples were then sputtered lightly to reduce their contribution and scanned again. Pre-sputtering resulted in increased amount of oxygen, and reduced carbon in each XPS analysis, which indicates that the nucleated surfaces get oxidized readily following post-experiment exposure to atmospheric conditions. The tin was always present as an oxide. The data obtained from samples with nucleation temperature above 400 °C showed (20-30)% total tin. A typical XPS scan is shown in Fig. 5. Note the absence of chlorine peaks: no chlorine was ever found in any of the nucleated samples. The amount of tin found with XPS on the surface is clearly more than we expected from the tin particles alone, which covered less than 10 % of the surface.

Table 2: XPS elemental analysis of nucleated samples at different temperature. Nucleation time for each sample was one hour.

| Nucleation Temperature (ºC) | Sample | C (at. %) | O (at. %) | Nb (at. %) | Sn (at. %) | $\frac{Sn}{Nb+Sn} \times 100$ | |
|---|---|---|---|---|---|---|---|
| 300 | U55 | 48.5 | 32.5 | 13.7 | 5.3 | 27.89 | |
| | | 6.1 | 61.5 | 28.5 | 3.9 | 12.03 | With pre-sputtering |
| 400 | U73 | 35.0 | 41.5 | 9.4 | 14.1 | 60.00 | |
| | | 4.9 | 57.8 | 25.1 | 12.2 | 32.71 | With pre-sputtering |
| 450 | U59 | 62.5 | 25.5 | 5.7 | 6.3 | 52.5 | |
| | U47 | 9.0 | 54.2 | 28.6 | 8.2 | 22.28 | With pre-sputtering |
| 500 | U90 | 56.8 | 31.2 | 7.3 | 4.7 | 39.17 | |
| | | 13.3 | 52.5 | 25.3 | 8.9 | 26.02 | With pre-sputtering |

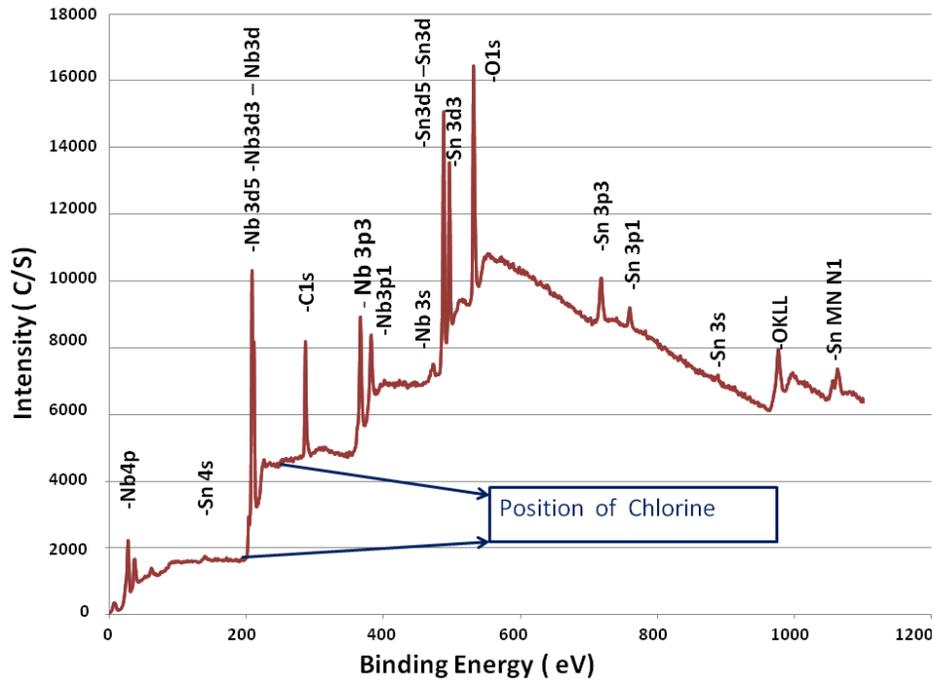

Figure 5: XPS data obtained from a sample after one hour at 500 ºC.

The discrepancy of tin content between SEM and XPS analysis indicated the presence of more tin other than visible tin particles in SEM. It can be explained by a thin tin layer on the surface. The layer is too thin to be revealed with SEM/EDS, but can be detected by XPS. A rough intensity ratio of Sn/Nb in XPS was then estimated considering a homogeneous Sn layer of monoatomic thickness in between tin particles on the niobium surface. For the case when there is a thin layer of Sn on top of bulk niobium, the layer and substrate intensities are given by [31],

$$I_{Sn} = I_{Sn}^{\infty} \left[1 - \exp\left(-\frac{d}{\lambda_{Sn,Sn} \cos\theta}\right)\right] \quad (1)$$

$$I_{Nb} = I_{Nb}^{\infty} \exp\left(-\frac{d}{\lambda_{Nb,Sn} \cos\theta}\right) \quad (2)$$

Where $I_{Sn}^{\infty}$ and $I_{Nb}^{\infty}$ are the intensities of the pure bulk tin and niobium. $\lambda_{Sn,Sn}$ and $\lambda_{Nb,Sn}$ are the attenuation length (inelastic mean free path) of the electrons in tin emitted by first subscript element to second subscript element. Emission angle, θ is the angle between the surface normal and direction of measured electron emission.
Dividing equation (1) by (2),

$$\frac{I_{Sn}}{I_{Nb}} = \frac{I_{Sn}^{\infty}\left[1 - \exp\left(-\frac{d}{\lambda_{Sn,Sn} \cos\theta}\right)\right]}{I_{Nb}^{\infty} \exp\left(-\frac{d}{\lambda_{Nb,Sn} \cos\theta}\right)} \quad (3)$$

The intensity ratio of bulk tin and niobium is given by [32],

$$\frac{I_{Sn}^{\infty}}{I_{Nb}^{\infty}} = \frac{N_{Sn}\sigma_{Sn}\lambda_{Sn}T_{Sn}}{N_{Nb}\sigma_{Nb}\lambda_{Nb}T_{Nb}} \quad (4)$$

Where N is atomic density, σ is photo-ionization cross-section for observed photoelectron line and T is transmission factor of instrument. From (3) and (4),

$$\frac{\frac{I_{Sn}}{\sigma_{Sn}\lambda_{Sn}T_{Sn}}}{\frac{I_{Nb}}{N_{Nb}\sigma_{Nb}\lambda_{Nb}T_{Nb}}} = \frac{N_{Sn}\left[1 - \exp\left(-\frac{d}{\lambda_{Sn,Sn} \cos\theta}\right)\right]}{N_{Nb} \exp\left(-\frac{d}{\lambda_{Nb,Sn} \cos\theta}\right)} \quad (5)$$

The ratio on the left-hand side gives the corrected intensity ratio, that is, the ratio of tin to niobium,

$$\frac{\eta_{Sn}}{\eta_{Nb}} = \frac{\frac{I_{Sn}}{\sigma_{Sn}\lambda_{Sn}T_{Sn}}}{\frac{I_{Nb}}{N_{Nb}\sigma_{Nb}\lambda_{Nb}T_{Nb}}} = \frac{N_{Sn}\left[1 - \exp\left(-\frac{d}{\lambda_{Sn,Sn} \cos\theta}\right)\right]}{N_{Nb} \exp\left(-\frac{d}{\lambda_{Nb,Sn} \cos\theta}\right)} \quad (6)$$

Let us use $3d_{5/2}$ photoelectron line for calculation. For E (Al $K_\alpha$) = 1486.7 eV,

K.E. $(3d_{5/2})_{Sn}$ = E (Al $K_\alpha$) – $E_b$ = 1486.7 – 484.9 = 1001.8 eV   (7)

Where K.E.$(3d_{5/2})$ and $E_b$ are the kinetic energy and binding energy of emitted $3d_{5/2}$ photoelectron.

For simplicity, $\lambda_{Sn,Sn} = \lambda_{Nb,Sn} = \lambda_{Sn}$, which can be extracted from NIST electron inelastic mean free path (IMFP) database, [33], using kinetic the energy of emitted photoelectrons. The required parameters to calculate the ratio of tin to niobium in equation 6 are tabulated here. θ = 45°, the one used in XPS measurement was chosen.

| $N_{Sn}$ | $N_{Nb}$ | $\lambda_{Sn}$ ($3d_{5/2}$) | Monolayer thickness (d) |
|---|---|---|---|
| 3.708 x $10^{22}$ | 5.55 x $10^{22}$ | 2.316 nm | 0.29 nm |

From SEM data, less than one tenth of total surface area appears to be covered with tin particles thick enough to shield all the photoelectrons from underlying niobium. The representative Sn/(Sn+Nb) ratio for a surface was found to be ~22%, obtained by taking the weighted average of contributions from Sn particles (10%) and monolayer of Sn on bulk Nb (90%). This ratio is consistent with Sn concentration range of XPS analysis, which supports the notion of a thin Sn film on nucleated surfaces. To further corroborate the finding, angle resolved XPS (ARXPS) was attempted in few samples. XPS data were collected by varying the angle between XPS detector and sample surfaces. XPS intensity from a thin film is expected to vary depending on the angle, since the angle changes the effective information depth, see equation 1. The shallower angles increase signal contribution from the thin film. ARXPS results are summarized in Table 3.

While all the four samples were nucleated at 500 ºC, M46 and U90 were subjected to standard protocol (1 hour of nucleation time with 1 g of $SnCl_2$). Two other samples, U101 and U66 were prepared with 5 mg (~ 10 µg/cm2) of $SnCl_2$ and 5 hours of nucleation time. Detailed description of preparation and analysis of similar samples will be presented later here. U66 and U90 demonstrated the expected change in intensity with change in angle for the presence of a continuous thin tin film. The other two studied samples, M46 and U101 did not show the similar trend. We propose that the pre-sputtering may have removed the shallow thin film on those samples. Lower ratio of Sn to Nb for these samples is consistent with this hypothesis.

Table 3: ARXPS ratio of Sn to (Nb+Sn).

| Samples | Angle between the surface normal and direction of electron emission (degree) | | | |
| --- | --- | --- | --- | --- |
| | 30 | 45 | 60 | 75 |
| U101 (~5mg $SnCl_2$) | 0.15 | 0.15 | 0.14 | 0.12 |
| M46 (1g $SnCl_2$) | 0.19 | 0.18 | 0.22 | 0.19 |
| U66 (~5mg $SnCl_2$) | 0.31 | 0.35 | 0.42 | 0.50 |
| U90 (1g $SnCl_2$) | 0.16 | 0.18 | 0.20 | 0.22 |

Furthermore, a few samples were examined with a scanning auger microscope (SAM). A 10 kV electron beam was used to bombard the sample surface producing Auger, secondary and backscattered electrons. Auger electrons are used to identify the elements present in the surface, whereas secondary and back scattered electrons are used for imaging at the same time providing an opportunity for surface sensitive elemental mapping in a SEM image.

Elemental mapping of Sn from a sample nucleated at 500°C for one hour is shown in Figure 6 (left). The observed Sn coverage was consistent with the XPS result. More Sn existed on the surface in between the particles visible in SEM image. Background between Sn particles was found even more Sn enriched than observed with XPS analysis in some cases. One such example of elemental comparison of Sn particle and background is also shown in Figure 6.

Transmission electron microscopy (TEM) imaging of the cross section of nucleated sample, treated at 500 ºC for 5 hours was done. The sectioning area was chosen with SEM, which contained particles as well as areas away from particles, representing an usual nucleated sample. EDS line scan was performed along the line, passing through and away from the particle in the cross-section. EDS line scan passing away from tin particle (see Figure 7) showed a brief jump of tin and oxygen signal close to the surface. This result is consistent with the previous finding of tin film by other characterization techniques discussed above.

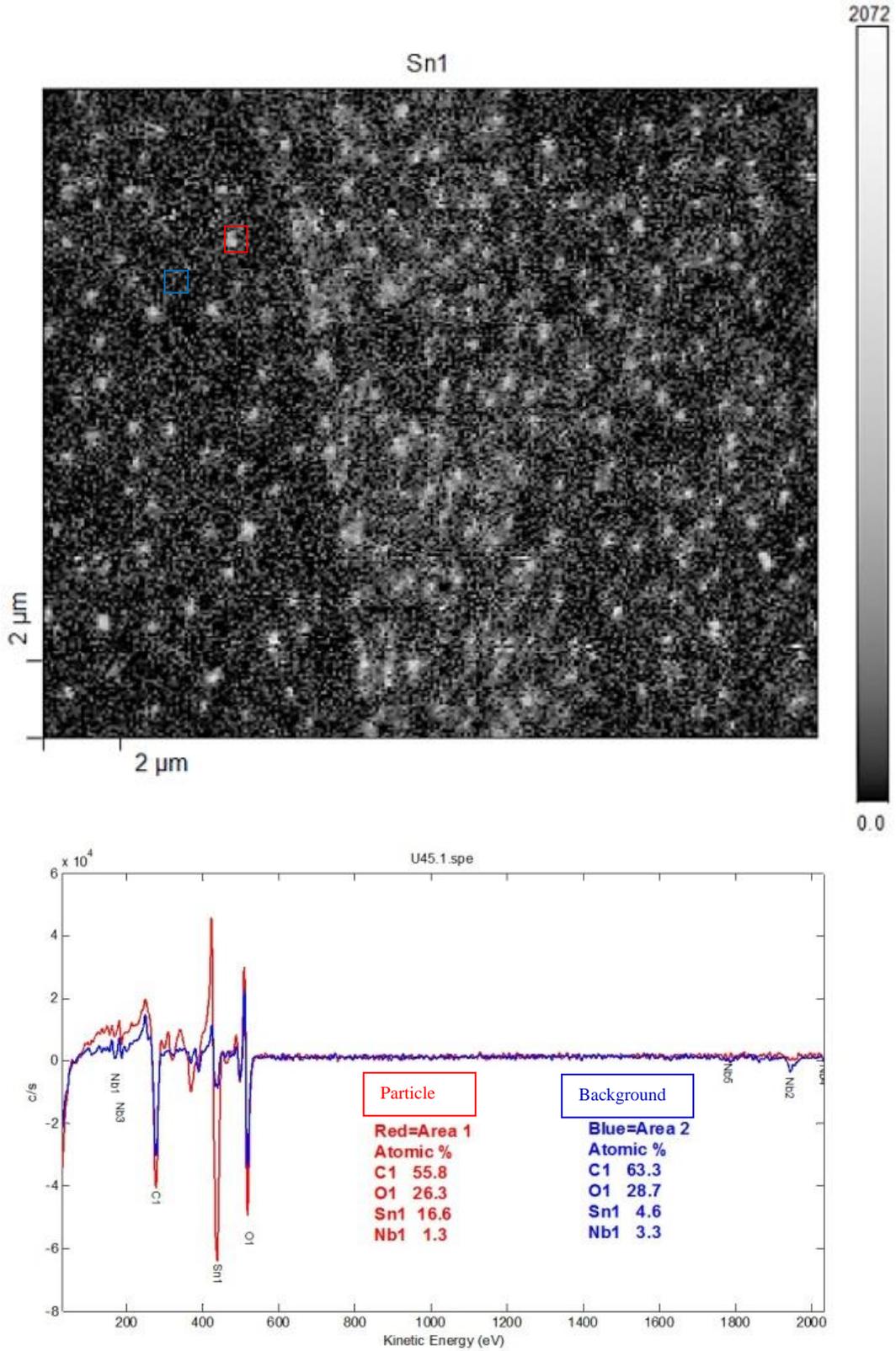

Figure 6: SAM elemental mapping of Sn coverage after sputtering for one minute is shown in the image (top). The brighter areas are richer in Sn than darker area as shown in intensity scale. Elemental composition comparison of particle and background is shown in spectra in the image at the bottom.

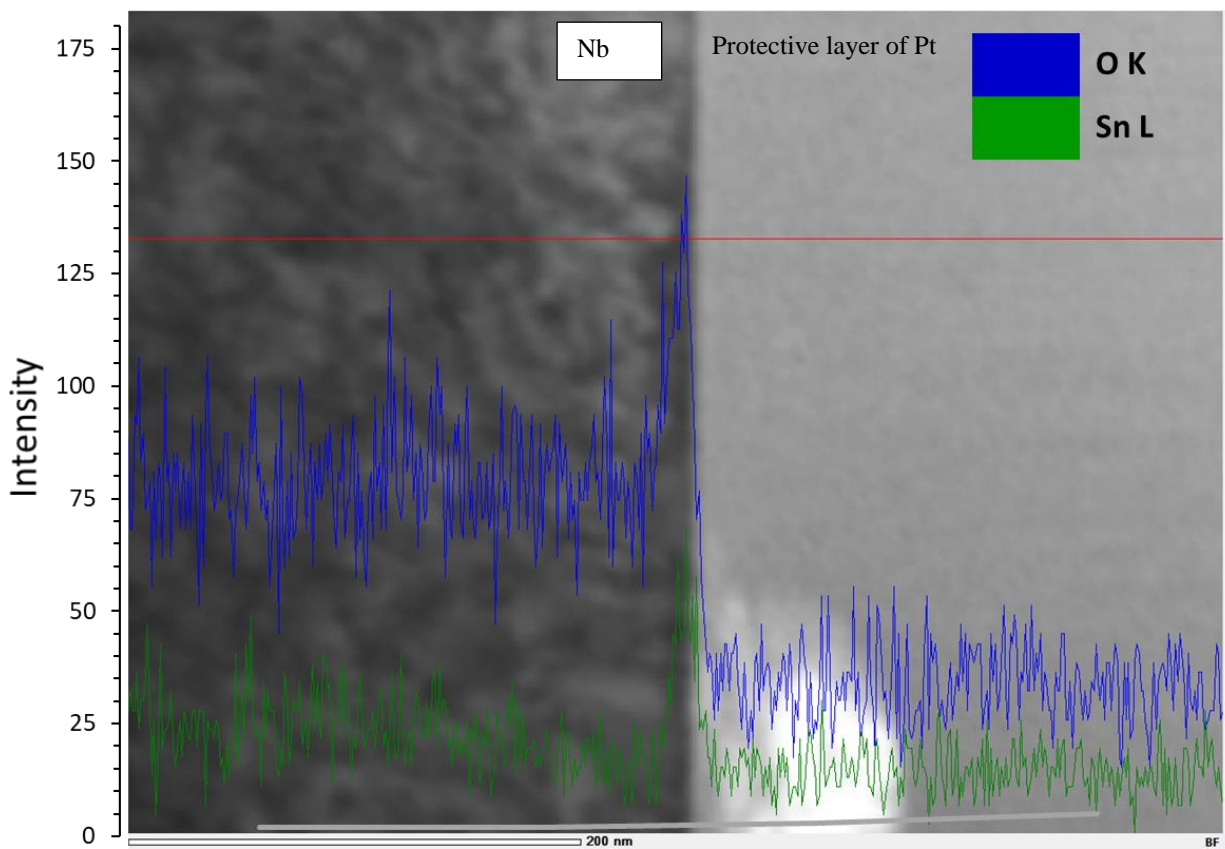

Figure 7: TEM cross section of sample nucleated at 500 ºC for 5 hours. EDS spectra from a line scan that follows solid red line is superimposed on top of TEM image. Nb signal is left out to make scaling reasonable for other signals. Note the jump of the Sn and O signal near the surface.

An AFM image from sample prepared under similar conditions, Fig. 8 (a) showed the evolution of tin particles as islands. More features, see area enclosed by rectangle in Fig. 8 (a) were seen in between tin islands. These features were developed after the experiment, as a typical nanopolished sample does not have such features. A height profile of a section from Fig 8 (a), passing through a big particle is shown in Fig 8(b). The diameter of the largest particle in this scan was ~200 nm with height of ~ 60 nm. Note that the largest particles observed were ~300 nm in diameter.

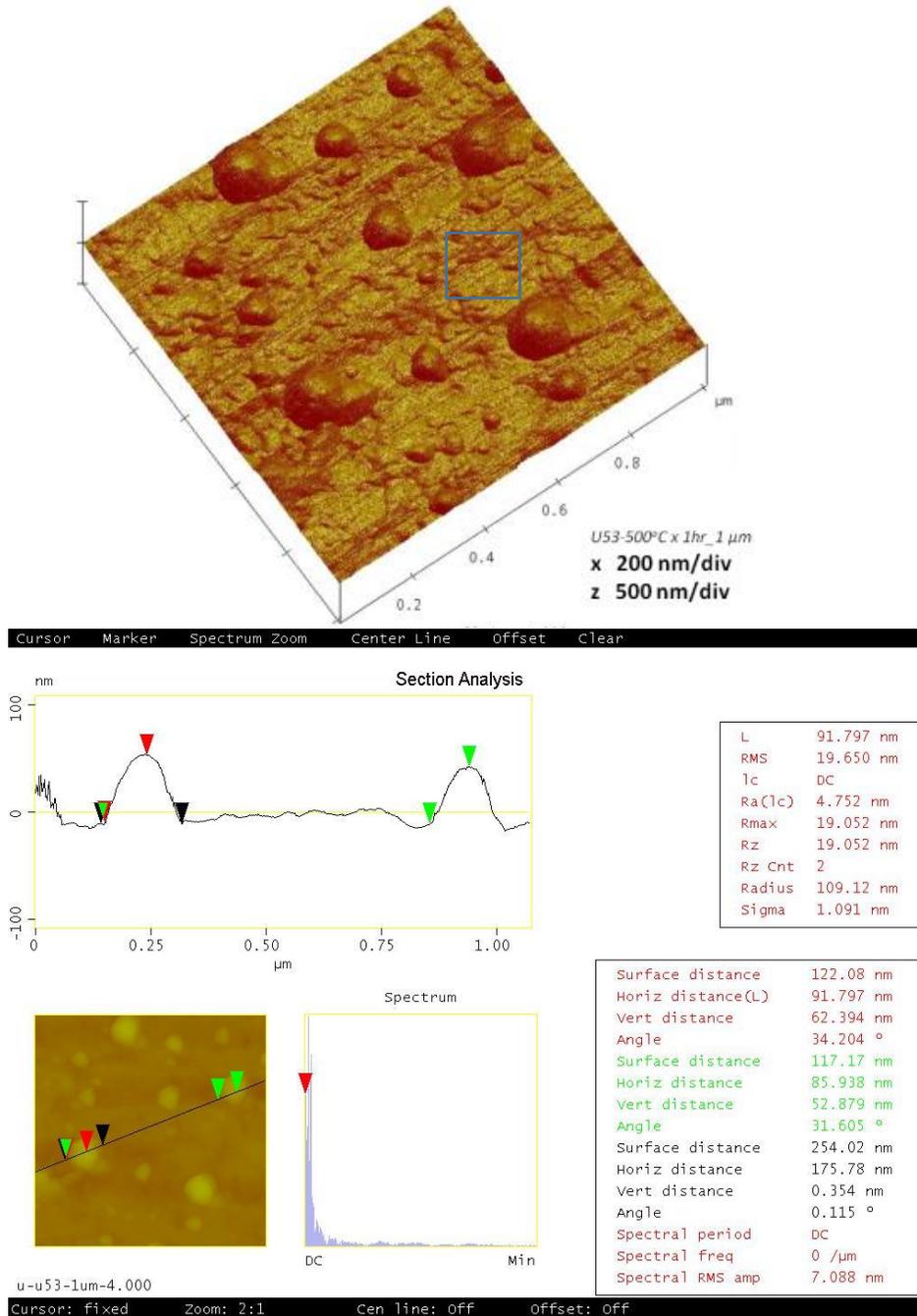

Figure 8: AFM image obtained from sample treated at 500 ºC for an hour is shown in (a). Note that scan size is only 1 µm x 1 µm. Sn particles appears to be three dimensional clusters forming islands. Height profile along the line passing through tin island is shown in (b). Triangles of the matching color are reference points for the height measurements.

Initial growth mode of thin film is commonly divided into three categories (Frank-Van Der Merve: two-dimensional (2D) layer growth; Stranski-Krastanov: layer plus three dimensional (3D) islands and Wolmer-Weber: 3D island growth). Our data qualitatively resembles Stranski-Krastanov mode; tin film is formed in addition to distributed tin particles. Similar growth has been reported before during the growth of tin on Al or Nb [34, 35]. It is important to note that the formation of a thin film, i.e., Frank-Van Der Merve growth, during nucleation was suggested from the early days of $Nb_3Sn$ diffusion coating [22, 36] . *The growth of such continuous layer has been suggested to be crucial to establish uniform coating of $Nb_3Sn$*, and the role of $SnCl_2$ and the anodic oxide layer was to retard Nb-Sn reaction until a uniform tin film is formed. However, previous studies of the nucleated surface only indicated the formation of tin particles like in Volmer-Weber mode on the surface following nucleation [15] , and the presence of a tin film was not established.

## 3.2  Nucleation time variation at 500°C

The nucleation period was varied by keeping the temperature constant at 500°C. It was found that five minutes at 500°C was not enough to evaporate all the Sn chloride (1 g), but it was able to produce some Sn particles already. The surface was covered with 'curly' features as shown in Figure 9, image to the right. Results obtained after an hour at the same temperature were discussed already in the previous section. The surface produced after four hours at 500 °C is also shown in Figure 9. Comparing with the result obtained after an hour at 500 °C, four hours at same temperature appears to produce additional small Sn particles resulting in more coverage on the surface. EDS was only able to detect Sn from visible Sn particles, but once again XPS analysis showed more Sn than SEM/EDS analysis reported. Comparison of XPS analysis is presented in Table 4.

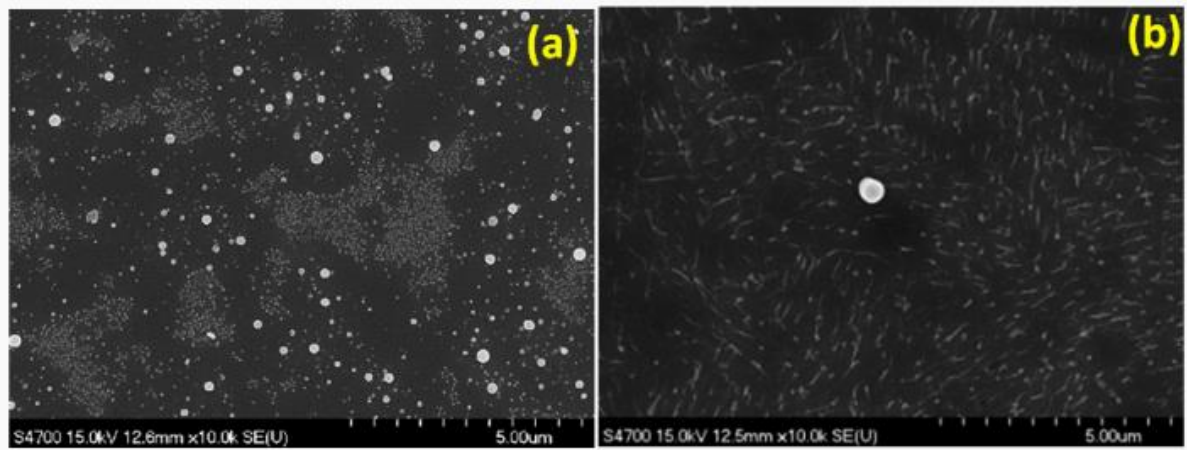

Figure 9: SEM images from samples obtained from experiments with nucleation temperature of 500 °C for (a) 4 hours and (b) 5 minutes.

Table 4: XPS elemental analysis of nucleated samples for different durations at 500 °C.

| Nucleation Temperature (°C) | Time | Sample | C (at. %) | O (at. %) | Nb (at. %) | Sn (at. %) | $\frac{Sn}{Nb+Sn} \times 100$ | |
|---|---|---|---|---|---|---|---|---|
| 500 °C | 4 hr | U1 | 7.2 | 61.3 | 23.9 | 12.0 | 33.42 | With pre-sputtering |
| 500 | 1 hr | U90 | 56.8 | 31.2 | 7.3 | 4.7 | 39.17 | |
| | | | 13.3 | 52.5 | 25.3 | 8.9 | 26.02 | With pre-sputtering |
| 500 °C | 5 min | U9 | 39.9 | 42.0 | 11.9 | 6.2 | 34.25 | |
| | | | 2.4 | 42.7 | 51.3 | 3.6 | 6.55 | With pre-sputtering |

AFM image captured from a sample prepared at the nucleation temperature of 500 °C for 4 hours is shown in Fig. 10 (a). Note that the size of the largest tin particle (see height profile of a section at Fig. 10 (b)) is similar at the base when compared to that obtained after an hour at 500 °C (Fig. 8), but elevated.

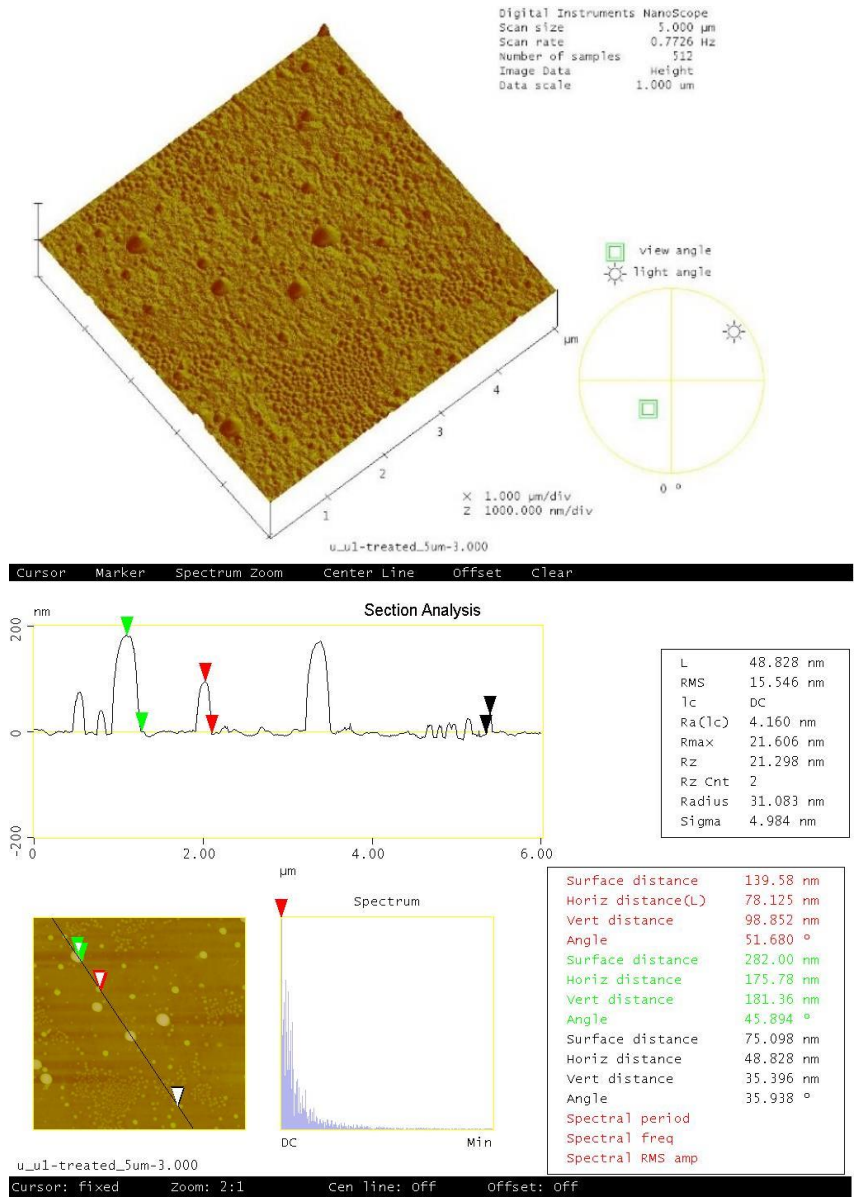

Figure 10: AFM image of the surface after 4 hours of nucleation at 500 °C is shown in (a). (b) shows height profile along a line passing through tin particles. The largest particle is ~ 300nm in diameter with height of ~175 nm. Triangles of the matching color are reference points for the height measurements.

### *3.3 Low vs. high amount of tin chloride*

The recipe for diffusion coating of $Nb_3Sn$ first developed at Siemens AG used a very small amount of tin halide (20 μg/cm$^2$). Recent recipes in practice at different research facilities use different amounts of tin chloride. We compare the nucleated surfaces produced by using a regular nucleation protocol at Jlab (~3 mg/cm$^2$ of tin chloride, 500°C for an hour, temperature ramp at 6 °C/min) and another protocol which uses much less tin chloride, reported by Cornell University (~10 μg/cm$^2$ of tin chloride, 500°C for five hours, temperature ramp at 3 °C/min) [37], as shown in Figure 11. SEM images show uniform distribution of particles produced with low amount of tin chloride. They are very small compared to the nucleation produced with a higher amount of tin chloride. EDS examination showed only Nb indicating that these particles are thin. XPS analysis, Table 5 shows comparable tin coverage in nucleated samples obtained from both experiments. It is clear that more tin chloride produces bigger tin particles with similar tin coverage on niobium surface. The longer nucleation time in later recipe is speculated to have an important role to produce such a good coverage of tin with a small amount of tin chloride.

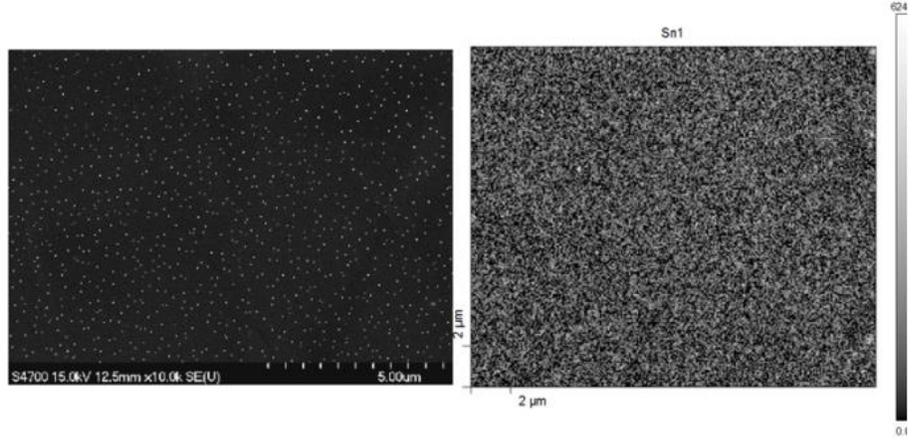

Figure 11: (a) SEM image of nucleated surface obtained by using low tin chloride. (b) SAM elemental mapping of tin.

Table 5: XPS elemental analysis of nucleated samples for different durations at 500 °C.

| Nucleation Temperature (ºC) | Time | Sample | C (at. %) | O (at. %) | Nb (at. %) | Sn (at. %) | $\frac{Sn}{Nb+Sn} x100$ | |
|---|---|---|---|---|---|---|---|---|
| 500 | 5 hr | U66 | 30.5 | 40.7 | 11.1 | 17.7 | 61.45 | |
| | | | 6.0 | 45.5 | 39.7 | 8.8 | 18.1 | with pre-sputtering |

### 3.4 Nucleation on anodized surfaces

Substrate anodization was introduced in 1970's to overcome non-uniformity and was often combined with the setup that maintains higher temperature of tin source compared to the substrate [22]. Despite the expected RRR loss, recent coating experiments with anodized substrates using the "standard" nucleation protocol indicated a positive effect for the coating uniformity [38, 39]. The effect of tin chloride in anodized and non-anodized niobium surfaces following the nucleation step was compared. A fixed cell voltage of 30 V or 50 V was applied to grow anodic oxide layers on BCP treated Nb samples with 15% $NH_4OH$ solution as an electrolyte. Thickness of the oxide later was estimated to be 72 nm and 120 nm for the samples using the thickness-voltage ratio [40].These samples were subjected to 5 hours of nucleation at 500 °C with the usual amount of Sn/$SnCl_2$. SEM images of the obtained surfaces are shown in Figure 12. Bright features were present at the surface following nucleation in each sample, but the appearance and distribution of these features (presumably tin) were different from observed for non-anodized samples. Bright features were bigger and sparsely distributed in a 50 V anodized sample compared to those in 30 V anodized sample. Our result appears to be different from recent results from similar studies [41] , which report the formation of big tin particles on the surface after pre-anodization unlike non-anodized Nb. Note that, smaller amount of tin chloride and different anodization parameters were used in latter studies compared to our experiments.

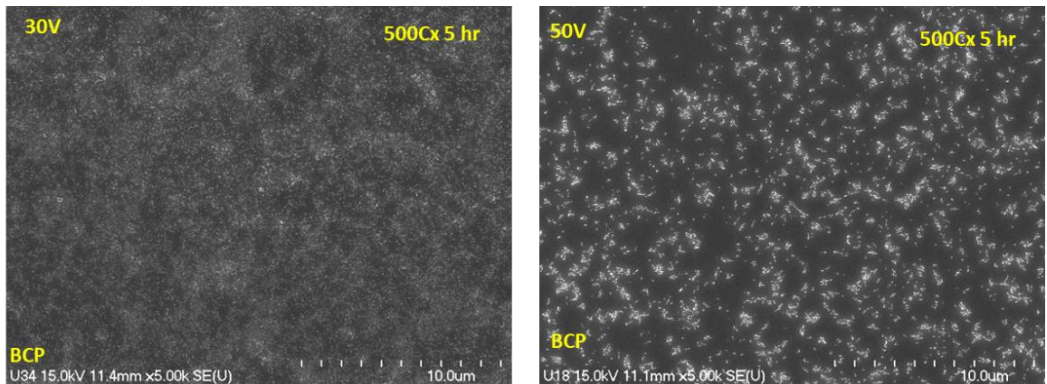

Figure 12: Nucleation on anodized niobium.

## 3.5 Variation of tin particle density

During the course of nucleation studies, the accumulation of visible tin particles seemed to vary for different grains of niobium in some cases as shown in Figure 13. The size and shape of tin particles on different niobium grains also varied in this case. Area A1 appears to have fewer particles that appear to be near circular under SEM whereas area A2 seems to have more particles with noncircular shape. Some surface defects in substrate niobium seemed to be the favorable location for particle accumulation in many cases. One example of such defect is a surface scratch, as shown in Figure 13, developed during the polishing of niobium sample. Each scratch line is decorated with a higher density of particles following the nucleation.

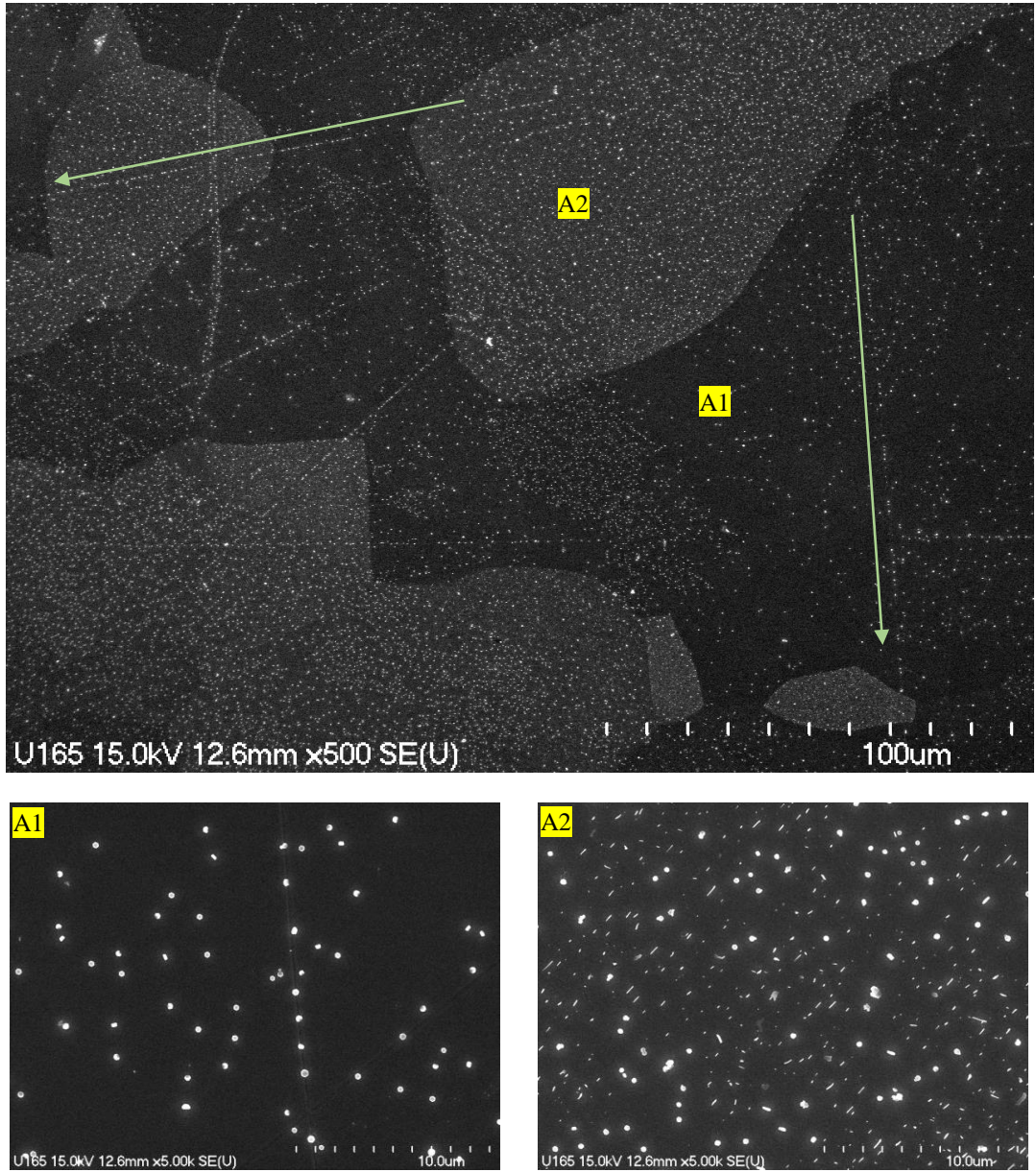

Figure 8: Variation of tin particle in different Nb grains.

## 3.6 Complete Nb$_3$Sn coatings with different nucleation profiles

A selected few nucleation profiles were then used for complete Nb$_3$Sn coating. The temperature profile used during the coating was similar to Figure 2, except for the variation in nucleation parameters. There were no evident differences in terms of structure and composition in SEM/EDS between the coatings produced with different nucleation profiles:

400 °C for an hour, 500 °C for an hour and 500 °C for 4 hours for similar amount (3 mg/cm$^2$) of tin and tin chloride. Obtained SEM images are shown in Figure 14, which also includes the coating produced by applying relatively very low amount of tin chloride for nucleation.

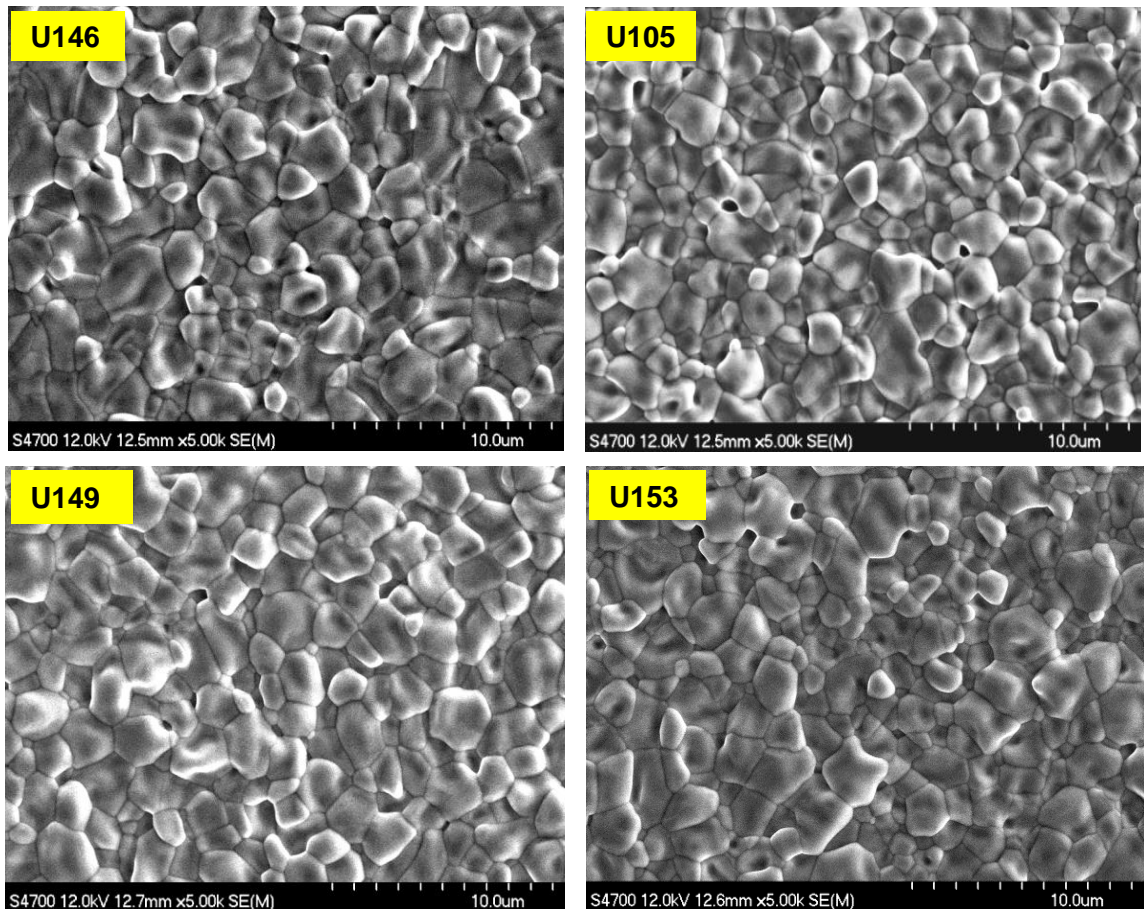

Figure 9: Nb$_3$Sn coatings obtained by applying different nucleation profiles. U146, U105, U149 and U153 involved nucleation step 400°C x 1hour, 500°C x 1hr and 500°C x 4 hour and 500°C x 5 hours respectively. U153 was coated with only ~ 5 mg of SnCl$_2$, and involved additional coating time of 6.5 hours at 1100°C.

## 3.7 Role of nucleation in coating genesis

The coating process was interrupted at different stages with or without tin chloride, while following the standard temperature profile as shown in Figure 2. The heat was turned off after 1 or 5 minutes, 1 hour and after the complete 3 hours. SEM images obtained from each experiment are presented in Figure 15 and 16. A minute after reaching at 1200 °C, a uniform coating with grain sizes of few tens of nanometers was developed on the whole niobium surface, when coated with tin chloride. Some tin droplets were visible. Similar experiments for 5 minutes at 1200 °C without tin chloride resulted in coating that included many patchy areas with irregular grain structures. These patchy areas appeared both after one hour, and the complete coating for 3 hours at 1200 °C. On the other hand, coatings obtained from similar experiments with provided tin chloride demonstrated little to no patchy regions. Repeated complete coating experiments without tin chloride often produced patchy regions with irregular grain structure. This indicated that the inclusion of nucleation step with tin chloride helps to assure a uniform Nb$_3$Sn coating compared to that without tin chloride. Note that patch-free coatings were also obtained in an experiment without tin chloride. This may have been due to the higher tin evaporation rate than normal, since the supplied tin was not packaged in niobium foil, thus providing relatively more surface area for evaporation.

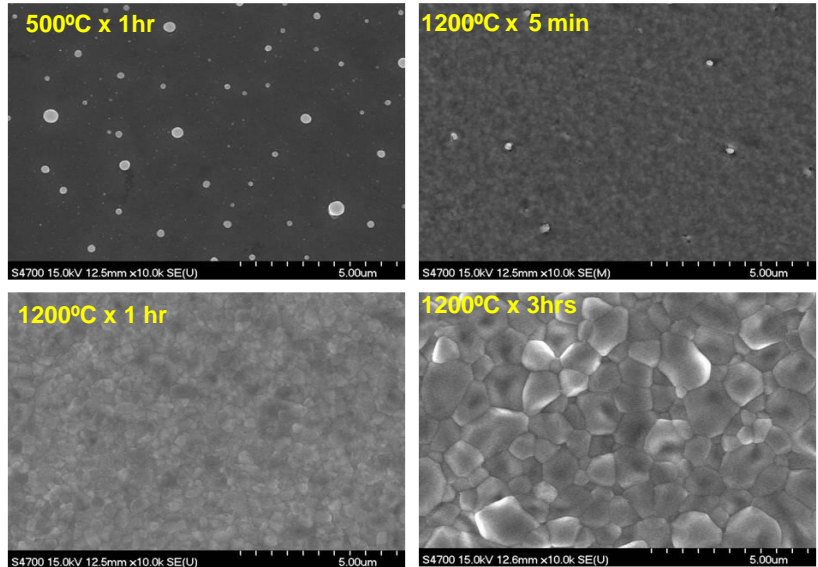

Figure 15: Different stages of Nb$_3$Sn coating evolution during the coating process (see Figure 1) including tin chloride. The bigger tin particles are still visible after 5 minutes at 1200°C.

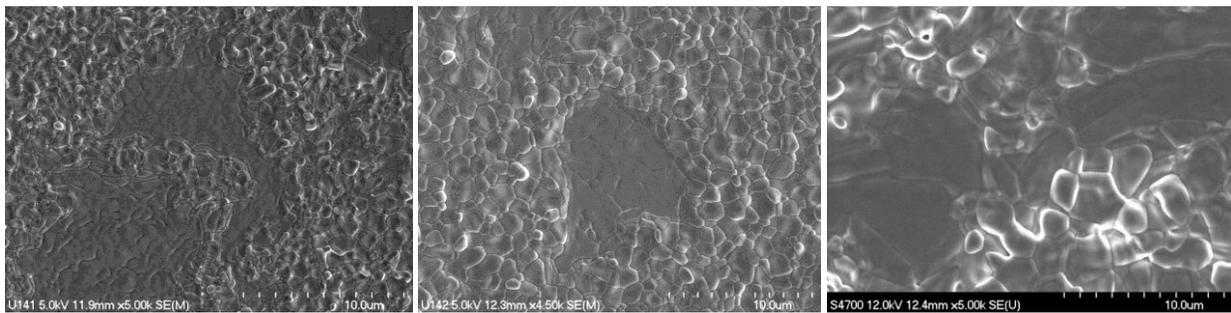

Figure 10: Different stages of Nb$_3$Sn coating evolution during the coating process excluding tin chloride Coatings obtained after 1 minute, 1 hour and 3 hours at 1200 °C are shown respectively.

Finally, we attempted to grow the complete coating without any tin, i.e., only SnCl$_2$ was used. The coating process included the nucleation (500 °C for 5 hours) and deposition (1200 °C for 3 hours) profile with 2 g of tin chloride. SEM image of the coating is shown in Figure 17. Complete coverage of coating was observed with some patchy areas and elongated grains. However, it shows that SnCl$_2$ plays an important role to establish a continuous tin layer prior tin evaporation during the coating process.

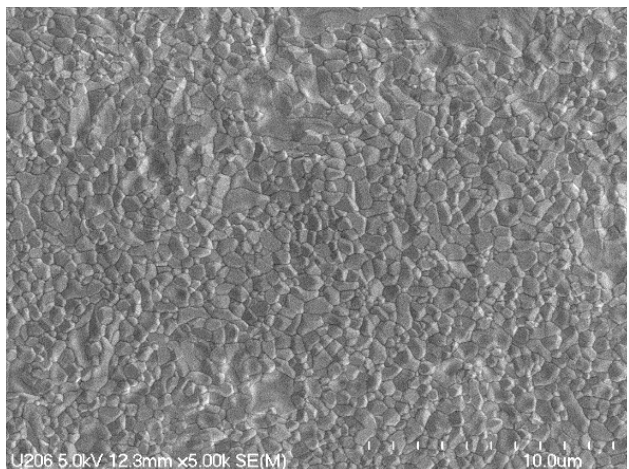

Figure 17: Coating prepared with only SnCl$_2$.

### *3.8 Mechanism of tin deposition via SnCl$_2$*

The as-prepared niobium surface is naturally covered with a 4-6 nm thick oxide layer because of high chemical reactivity [42] . Nb$_2$O$_5$ is the dominant form of oxide because of its higher free energy of dissociation compared to other oxide forms. The presence of other suboxides (NbO$_2$, NbO, Nb$_2$O, Nb$_2$O$_3$) was inferred from Nb binding energy observed by XPS [43, 44] . The thermal stability of native oxide layer has been studied with respect to SRF cavities. These works generally agree that the typical Nb$_2$O$_5$ layer reduces to NbO$_2$, and then to NbO following UHV baking at temperature >300 °C [43] . The experimented nucleation temperatures should be in the range where the reduction of native Nb$_2$O$_5$ layer progresses.

The deposition of tin-rich features at the surface following each experiment involve some sort of interchange or reduction reaction with SnCl$_2$, which may happen in three different ways in our case. First, SnCl$_2$ can react with some small amount of hydroxyl present in the nucleation chamber. Second, the possible interaction with niobium oxide covering the surface. Another possibility is direct reaction between SnCl$_2$ and niobium surface. Representative possible reactions are

$$SnCl_2 + H_2 \rightarrow Sn + 2HCl \quad\quad (a)$$

$$3SnCl_2 + 2NbO \rightarrow 2\,NbOCl_3 + 3Sn \quad\quad (b)$$

$$x\,SnCl_2 + 2\,Nb \rightarrow 2\,NbCl_x + x\,Sn \quad\quad (c)$$

Gibbs free energy data, [45] shows that reaction (a) happens at ~1500 °C under standard conditions, which indicates that it is less favoured thermo-chemically in our experimented nucleation temperatures. On the other hand, reduction of SnCl$_2$ with hydrogen had been reported before to deposit Sn on copper in similar scenario, where they mention that it is necessary to keep HCl pressure at low level by removing it for such reaction [46]. We can expect a low level of HCl in experiment as it was constantly pumped during the experiment. As both hydrogen and HCl were observed with residual gas analyser (RGA) during some of our nucleation experiments, we believe that the tin deposition during the nucleation process is contributed by reduction of SnCl$_2$ with already available hydrogen in the reaction chamber. Since, possible chlorination of Nb$_2$O$_5$ at low temperature through an intermediate step involving the formation of niobium oxychloride has been suggested before [47-49], we consider the possible interaction of SnCl$_2$ with niobium oxide. XRR studies on Nb(110) claim that Nb$_2$O$_5$ and NbO$_2$ both reduce to NbO following 30 minutes of vacuum annealing at 300 °C [50] . So, we only illustrated equation (b) here, but the thermodynamic data are not all available to check its feasibility. In reaction (c), x can be 3, 4 or 5. It is found that NbCl$_3$ is favored the most for any temperature < 500 °C compared to NbCl$_4$ under standard condition. NbCl$_5$ is not thermodynamically favored in our experimental temperatures [45]. Besides the possible reactions producing tin, concomitant reduction of niobium oxide layer upon the arrival of SnCl$_2$ vapor, the mobility and surface diffusion of tin, niobium surface properties (e.g. defects, orientation) are other factors to be considered to explain the produced nucleated surfaces from different experiments.

Since the evaporation of SnCl$_2$ is expected to start at ~ 250 °C, the known transformation of the native Nb2O5 surface in this temperature range occurs in its presence. We assume that as we heat the niobium surface, oxide dissolution results in randomly distributed defects first at lower temperature. These sites are favorable to trap tin early. Because of low amount of SnCl$_2$ evaporation at lower temperature (~ 300 °C), these particles are only of few nanometers in size. Further heating increases both the defect population and the SnCl$_2$ partial pressure. Such locations enhance SnCl$_2$ interaction with the Nb surface producing more tin, bringing in features shown in Figure 4.b at about 400 °C. Raising nucleation temperature further to (450-500) °C is expected to produce more defects in oxide layer leading to more particles. The native niobium oxide layer would have reduced completely after few minutes at these temperatures. A reaction between SnCl$_2$ and niobium may take place uniformly, as suggested by reaction (c) at this point forming tin film by the direct reaction between SnCl$_2$ and Nb. Longer nucleation may add further tin to the thin film and particles. The equilibrium surface composition is determined by tin arrival as well as tin-tin and tin-surface interaction.

Surface defects, like the scratch discussed here have different surface diffusion rates from flat surface [51]. Sn atoms bond well to these less coordinated sites, and they serve as sinks for Sn during Sn vapor deposition favoring the formation of tin particles.

The different density of particles on different grains points to an orientation-dependent adatom-surface interaction. One can expect the anisotropy in tin adsorption because of anisotropy surface energy. For example Nb(110) has the lowest surface energy because of bcc structure [52].

## 4 Conclusions

Nucleation parameters were varied for experimental investigation of the nucleation process used in Nb$_3$Sn diffusion coating. Examination of several sample prepared under different conditions leads to the following conclusions

- No chlorine was detected after any experiment.
- SEM images obtained from samples nucleated at 450 °C and higher temperature show a distribution of the tin particles. AFM examination showed consistent three-dimensional island structures of tin particles.
- XPS and TEM analysis along with SAM elemental mapping reveal a clear presence of additional tin (film) in between the tin particles. It resembles surfaces after the Stranski-Krastanov growth mode, where tin film is formed in addition to distributed tin particles.
- Amount of tin chloride is found not to be crucial for surface coverage of tin on niobium surface, but a larger amount appears to create bigger tin particles. Longer nucleation time appears to create additional tin particles of relatively small size compared to shorter nucleation time.
- Even though the variation of nucleation parameters was able to produce drastically different surfaces following nucleation, no evidence of any significant impact on the final coating was found with SEM/EDS.
- Despite some patchy region and irregular grain structure, SnCl$_2$ alone was able to produce continuous coating, which showed importance of nucleation for coverage. The thin film of tin deposited during the nucleation appears crucial for such coverage.
- Three reactions are suggested as possible mechanisms for tin deposition onto niobium surface. The resulting tin density is affected by the interaction of tin adatoms with surface defects and by orientation-dependent adatom-surface interaction.
- The inclusion of the nucleation step appeared to be advantageous to produce uniform Nb$_3$Sn coating compared to the coating obtained without tin chloride. Absence of SnCl$_2$ increased the likelihood of forming patchy regions with irregular grain structure.

## 5 Acknowledgements

Thanks to Olga Trifimova for her help with AFM. SAM measurement was done at Swagelok Center for Surface Analysis of Materials at Case Western Reserve University. Work at College of William & Mary supported by Office of High Energy Physics under grant SC0014475. Partially authored by Jefferson Science Associates under contract no. DEAC0506OR23177.